\documentclass[conference]{IEEEtran}
\IEEEoverridecommandlockouts
\usepackage{booktabs}
\usepackage[utf8]{inputenc}
\usepackage{amsmath}
\usepackage{amsfonts}
\usepackage{amssymb}
\usepackage{graphicx}
\usepackage{listings}
\usepackage{algorithm}
\usepackage[noend]{algpseudocode}
\usepackage[utf8]{inputenc}
\usepackage[english]{babel}
\usepackage{xspace}
\usepackage{tabularx}
\usepackage{paralist}
\usepackage{multirow}
\usepackage[table,xcdraw]{xcolor}
\usepackage[T1]{fontenc}
\usepackage[scaled=.8]{beramono}
\def\BibTeX{{\rm B\kern-.05em{\sc i\kern-.025em b}\kern-.08em
    T\kern-.1667em\lower.7ex\hbox{E}\kern-.125emX}}

\newcommand{\tool}{\textsc{CoFL}\xspace}
\newcolumntype{L}[1]{>{\raggedright\arraybackslash}p{#1}}

\usepackage{amsthm}

\usepackage{xcolor}
\usepackage{hyperref}
\usepackage{enumitem}
\usepackage{balance}

\definecolor{dkgreen}{rgb}{0,0.6,0}
\definecolor{gray}{rgb}{0.5,0.5,0.5}
\definecolor{mauve}{rgb}{0.58,0,0.82}
\makeatletter
\renewcommand{\ALG@beginalgorithmic}{\small}
\makeatother
\begin{document}

\title{Configuration-dependent Fault Localization}

\author{\IEEEauthorblockN{Son Nguyen}
\IEEEauthorblockA{\textit{The University of Texas at Dallas} \\
800 W. Campbell, Richardson, TX 75080, USA \\
sonnguyen@utdallas.edu}}

\maketitle

\begin{abstract}
In a buggy configurable system, configuration-dependent bugs cause the
failures in only certain configurations due to unexpected interactions
among features. Manually localizing configuration-dependent faults in
configurable systems could be highly time-consuming due to their
complexity. However, the cause of configuration-dependent bugs is not
considered by existing automated fault localization techniques, which
are designed to localize bugs in non-configurable code. Thus, their
capacity for efficient configuration-dependent localization is
limited.
In this work, we propose {\tool}, a novel approach to localize
configuration-dependent bugs by identifying and analyzing suspicious
feature interactions that potentially cause the failures in buggy
configurable systems.
We evaluated the efficiency of {\tool} in fault localization of
artificial configuration-dependent faults in a highly-configurable
system. We found that {\tool} significantly improves the baseline
spectrum-based approaches. With {\tool}, on average, the correctness
in ranking the buggy statements increases more than 7 times, and the
search space is significantly narrowed down, about 15 times.

\end{abstract}

\section{Problem Statement and Background}
Configurable system supports the diversification of software
products by providing \textit{configuration options} that are 
used to control different \textit{features}. However, this 
induces challenges in program analyses and quality assurance 
~\cite{productlinesbook1, productlinesbook2, productline_survey}.

In quality assurance for configurable system,
\textit{configuration-dependent faults}, which cause the failures in
only certain \textit{configurations} because of unexpected
\textit{interactions} among several features, are not rare~
\cite{Garvin:2011, plugin_testing, Kuhn:2004, Yin:2011}.  Manually
localizing configuration-dependent faults in configurable systems
could be highly costly due to their complexity~\cite{Meinicke:2016,
  productline_survey}.

Meanwhile, existing automated fault localization techniques~\cite{wong2009survey} are designed to localize the faults in non-configurable code. Specifically, for configurable code, they do not consider the cause of configuration-dependent bug(s), which is the unexpected feature interactions. Thus, many parts of the buggy system, which are not related to those unexpected interactions, are inappropriately considered as suspicious. 
Indeed, for example, despite that one can adapt spectrum-based
techniques~\cite{tarantula, ochiai, wong2009survey} for configurable
code by considering static conditional statements (e.g.,
\texttt{\#if}) on configuration options as \texttt{if}-statements,
the adapted techniques still 
%spectrum-based techniques~\cite{tarantula, ochiai, wong2009survey}
access and rank all executed statements including
%based on the test execution information, and these statements might
the ones that might not affect the fault-inducing interactions, even
not the program's states.
%{\bf Son: don't understand!}
%
%By these techniques, the statements that have a similar execution 
%pattern are asigned the same suspiciousness assessment
%~\cite{Pearson:2017, exam, wong2009survey}.
%
For slice-based methods~\cite{static_slice, dynamic_slice}, the
suspicious domain is reduced to all slices that are related to failed
test execution information, which might include the slices irrelevant
to the unexpected feature interactions.
Therefore, the capacity of the traditional
techniques~\cite{wong2009survey} for efficient configuration-dependent
fault localization is limited.

%Though spectrum-based techniques such as Tarantula~\cite{tarantula} 
%can be used on configurable code by considering options 
%as conditions, these approaches access statements based on their 
%execution information and lack the consideration of the meaning of 
%statements in configurable code. Consequently, they provide the same 
%suspiciousness assessment to every statement that have similar 
%execution patterns~\cite{Pearson:2017}. For example, since options are 
%considered as program conditions, they might assign the same 
%suspiciousness level to all statements in the whole feature. That makes 
%configuration-dependent fault localizing becomes extremely inefficient~
%\cite{Pearson:2017, exam, wong2009survey}.

\section{Motivation and Observation}
\begin{figure}
\centering
\includegraphics[width=0.42\textwidth]{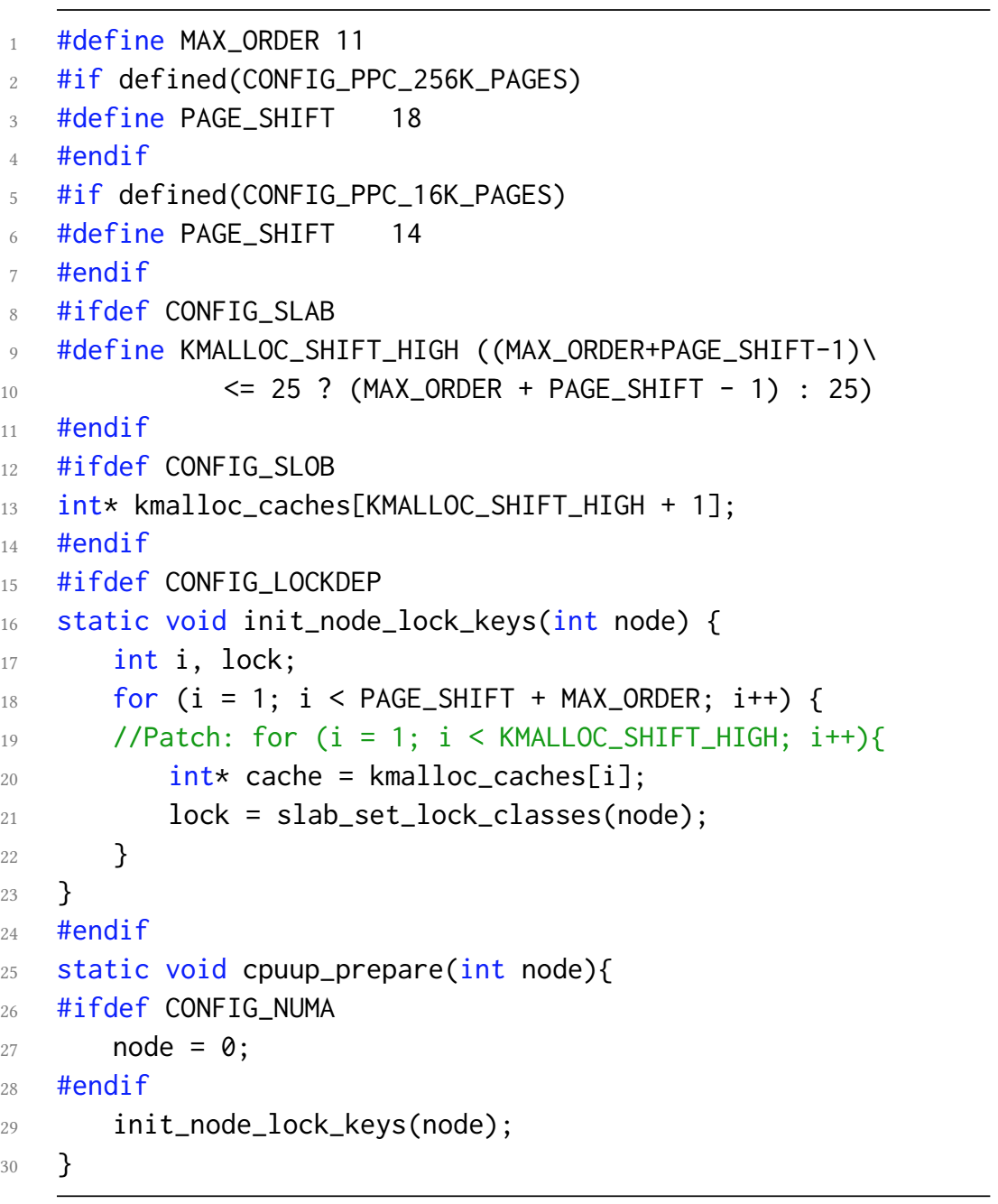}
\caption{A Configuration-dependent Bug in Linux Kernel}
\label{example_bug}
\end{figure}

Let us start with a real configuration-dependent bug in Linux kernel to motivate our approach (Fig.~\ref{example_bug}).
In this example, the maximum value of 
\texttt{KMALLOC\_SHIFT\_HIGH} is \texttt{25} (lines 9--10). This 
indicates that \texttt{kmalloc\_caches} contains a maximum of 26 
elements (line 13). When \texttt{PPC\_256K\_PAGES} is 
enabled and \texttt{PPC\_16K\_PAGES} is disabled, the maximum index used to access \texttt{kmalloc\_caches} is defined as
\texttt{(PAGE\_SHIFT + MAX\_ORDER-1)} (line 18), which is
\texttt{28}. This leads to an exception that array
\texttt{kmalloc\_caches} is accessed out of its bounds. However, this
bug is not revealed by any configuration, except the configurations in which \texttt{PPC\_256K\_PAGES}, \texttt{SLAB},
\texttt{LOCKDEP}, and \texttt{SLOB} are enabled, and
\texttt{PPC\_16K\_PAGES} is disabled.

\textbf{Observations.} From the example shown in Fig.~\ref{example_bug}, 
we have the following observations:

{\bf O1.} \textit{In a configurable system containing configuration-dependent bug, there are certain features that are (ir)relevant to the visibility of the bug}.  
%In Fig.~\ref{example_bug},
%since the system fails in only certain configurations, not all
%features are relevant to the visibility of the bug.
%
For example, in Fig.~\ref{example_bug}, feature \texttt{NUMA} (line 27) does not involve in the bug because when \texttt{PPC\_256K\_PAGES}, \texttt{SLAB}, \texttt{LOCKDEP}, and \texttt{SLOB} are enabled and \texttt{PPC\_16K\_PAGES} is disabled, the system still fails regardless of whether \texttt{NUMA} is enabled or disabled. Meanwhile, for some configurations, enabling/disabling certain features might make the test results (passing all tests or not) of the resulting configurations change. In Fig.~\ref{example_bug}, the all-enabled configuration behaves as expected, while if \texttt{PPC\_16K\_PAGES} is disabled and all other options enabled, the resulting configuration fails.
%re-phrase
%

{\bf O2.} \textit{In the features $f_E$s that \textbf{must be enabled} to make the bug visible, only the statements that implement the interaction between them are more likely to be buggy than others}.
In \texttt{LOCKDEP}, the buggy statement is at line 18, which is one
of the statements realizing the interaction between $f_E$s.
In~contrast, if the bug is caused by the statements not related to
the~interaction between $f_E$s, the visibility of the bug would not
depend on all of those $f_E$s.
In Fig. \ref{example_bug}, the enabled features $f_E$s include \texttt{PPC\_256K\_PAGES},\texttt{SLAB}, \texttt{LOCKDEP}, \texttt{SLOB}, and \texttt{PPC\_16K\_PAGES}. The bug is not related to the statement at line 21 in \texttt{LOCKDEP}, which is not used to realize the interaction of $f_E$s.
%
%In contrast,

%
{\bf O3.} \textit{In the features $f_D$s that \textbf{must be disabled} to make the bug visible, the statements that implement the interactions with $f_E$s also provide useful indication to help us find suspicious statements in $f_E$s}. In Fig.~\ref{example_bug}, \texttt{PPC\_16K\_PAGES} is a disabled feature $f_D$. Although line 6 in \texttt{PPC\_16K\_PAGES} (being disabled) is not considered as faulty, however analyzing the impact of the statement at this line (defining \texttt{PAGE\_SHIFT}) on the statements in \texttt{LOCKDEP} and \texttt{SLAB} can provide the suggestion to identify the statement need to be fixed (\texttt{i < PAGE\_SHIFT + MAX\_ORDER}).
The intuition of this phenomenon is that despite that the statements in $f_D$s are not faulty, $f_D$s have the impact of ``hiding''/``masking'' the bug when they are enabled. Thus, we need to consider the interactions of other features with $f_D$s in localizing configuration-dependent bugs.

{\bf O4.} Because certain statements in the enabled features~to make
the bug visible are considered as suspicious, the statements in the
same/different features having impacts on the suspicious
statements via program dependencies~\cite{cia,pdg} should also be
considered as suspicious. For example, although
line~1 does not belong to any $f_E$, that statement is also suspicious
since it has an impact on the statements at lines 9, 10, and~18.

\section{Approach}
We propose, {\tool}, a novel approach for configuration-dependent
fault localization. For a buggy configurable code, to reduce the
suspicious domain, \textbf{we analyze the test results of the executed
  configurations, the code, and the test execution information to
  identify the executed statements related to the interactions among
  the features whose enabling/disabling affect the visibility of the
  bugs which potentially cause the failures}. These statements are
ranked by their suspiciousness levels assigned by existing
techniques~\cite{wong2009survey} based on their test execution
information.

In particular, {\tool} first determines minimal sets of feature
candidates whose enabling/disabling (feature selection)
%\footnote{The
%  selection of feature $f$ is enabling/disabling ($f=T/F$ for short)})
make the bugs visible (based on \textbf{O1}). Let us call such a set
of feature selections the \textit{suspicious partial configuration
  (SPC)}. For example, \{\texttt{SLAB=T}, \texttt{PPC\_16K\_PAGES=F},
\texttt{PPC\_256K\_PAGES=T}, \texttt{LOCKDEP=T}, \texttt{SLOB=T}\} is
considered as the $SPC$ of the bug in Fig.\ref{example_bug}. The
selection of \texttt{NUMA} does not belong to the $SPC$ of the bug
because they do not have any impact on its visibility.

Next, {\tool} aims to detect the suspicious statements that are
responsible for the feature interactions and potentially cause the faults. To do that, it analyzes the features in $SPC$ to detect the interactions between them that are potentially cause/disguise the configuration-dependent bugs. Then, {\tool} detects the statements that
realize those interactions (based on \textbf{O2} and \textbf{O3}). The
interactions are detected via the shared program entities including \textit{variables} and \textit{functions} controlled by different features and the operations including \textit{define} and \textit{use} performed on them. For example, \texttt{PPC\_256K\_PAGES} \textit{define} \texttt{PAGE\_SHIFT} which is \textit{used} by \texttt{SLAB} and \texttt{LOCKDEP}. In the example, the statements realizing the interactions among the $f_E$s in the $SPC$ are at lines 3, 9, 10, 13, 18, and~20 ($S_1$). Meanwhile, the statements in $f_E$s for interactions between the $f_E$s and the $f_D$s in the $SPC$ are at lines 9 and 18 ($S_2$).

After that, the suspicious statements are used to detect other
suspicious statements that are executed and have dependencies on the
statements in both $S_1$ and $S_2$ in the failed configurations (based
on \textbf{O3} and \textbf{O4}). The output for the running example is
the set of statements at lines 3, 9, 10, 18, and 1. Finally, these
statements are ranked by their suspiciousness scores
computed by existing techniques~\cite{wong2009survey} such as
spectrum-based methods based on their test execution information.

\section{Empirical Evaluation}
We evaluate {\tool}'s efficiency in localizing configuration-dependent bugs over 2 spectrum-based techniques, Tarantula~\cite{tarantula} and Ochiai~\cite{ochiai}.
%, when we use their corresponding formula to compute the suspiciousness score.
We randomly seeded the set of 32 artifical configuration-dependent
bugs into the subject system BusyBox~\cite{busybox}. For each bug, the
output rank are evaluated via $EXAM$~\cite{exam} and the suspicious
domain size ($SDS$). The lower $EXAM$ and smaller $SDS$ the more
efficient the technique.

\begin{table}[!h]
\centering
\caption{Comparison Results}
\label{results_tb}
\begin{tabular}{|l|l|l|l|}
\hline
 						& $EXAM$	 	& $SDS$			\\ \hline
Tarantula 				& 37.50 		& 147.17			\\ \hline
{\tool} with Tarantula 	& 5.12  		& 10.58			\\ \hline
Ochiai 					& 36.54		& 147.17			\\ \hline
{\tool} with Ochiai 		& 4.97 		& 10.58			\\ \hline
\end{tabular}
\end{table}

Table \ref{results_tb} shows the average $EXAM$ and average $SDS$ of Tarantula, Ochiai and {\tool} with their formula. As seen, on average, the correctness in ranking the buggy statements increases more than 7 times, and the search space is significantly narrowed down, about 15 times.

\textbf{Conclusion.}
%We introduce {\tool}, an efficient configuration-dependent fault localization method for configurable code.
The novel idea of {\tool}, our configuration-dependent fault
localization method for configurable code, is to leverage the test
results and code analysis to detect interactions between features that
potentially cause the bugs and use these interactions to reduce the
suspicious domain. 

\balance
\bibliographystyle{plain}
\bibliography{references}

\begin{thebibliography}{10}

\bibitem{busybox}
Busybox: The swiss army knife of embedded linux, 2018.

\bibitem{ochiai}
Rui Abreu, Peter Zoeteweij, and Arjan J. C.~van Gemund.
\newblock An evaluation of similarity coefficients for software fault
  localization.
\newblock In {\em Proceedings of the 12th Pacific Rim International Symposium
  on Dependable Computing}, PRDC '06, pages 39--46, Washington, DC, USA, 2006.
  IEEE Computer Society.

\bibitem{dynamic_slice}
Hiralal Agrawal and Joseph~R. Horgan.
\newblock Dynamic program slicing.
\newblock In {\em Proceedings of the ACM SIGPLAN 1990 Conference on Programming
  Language Design and Implementation}, PLDI '90, pages 246--256, New York, NY,
  USA, 1990. ACM.

\bibitem{productlinesbook2}
Sven Apel, Don Batory, Christian Kstner, and Gunter Saake.
\newblock {\em Feature-Oriented Software Product Lines: Concepts and
  Implementation}.
\newblock Springer Publishing Company, Incorporated, 1st edition, 2016.

\bibitem{cia}
Robert~S. Arnold.
\newblock {\em Software Change Impact Analysis}.
\newblock IEEE Computer Society Press, Los Alamitos, CA, USA, 1996.

\bibitem{pdg}
Jeanne Ferrante, Karl~J. Ottenstein, and Joe~D. Warren.
\newblock The program dependence graph and its use in optimization.
\newblock {\em ACM Trans. Program. Lang. Syst.}, 9(3):319--349, July 1987.

\bibitem{Garvin:2011}
Brady~J. Garvin and Myra~B. Cohen.
\newblock Feature interaction faults revisited: An exploratory study.
\newblock In {\em Proceedings of the 2011 IEEE 22Nd International Symposium on
  Software Reliability Engineering}, ISSRE '11, pages 90--99, Washington, DC,
  USA, 2011. IEEE Computer Society.

\bibitem{plugin_testing}
Michaela Greiler, Arie~van Deursen, and Margaret-Anne Storey.
\newblock Test confessions: A study of testing practices for plug-in systems.
\newblock In {\em Proceedings of the 34th International Conference on Software
  Engineering}, ICSE '12, pages 244--254, Piscataway, NJ, USA, 2012. IEEE
  Press.

\bibitem{exam}
James~A. Jones and Mary~Jean Harrold.
\newblock Empirical evaluation of the tarantula automatic fault-localization
  technique.
\newblock In {\em Proceedings of the 20th IEEE/ACM International Conference on
  Automated Software Engineering}, ASE '05, pages 273--282, New York, NY, USA,
  2005. ACM.

\bibitem{tarantula}
James~A Jones, Mary~Jean Harrold, and John Stasko.
\newblock Visualization of test information to assist fault localization.
\newblock In {\em Software Engineering, 2002. ICSE 2002. Proceedings of the
  24rd International Conference on}, pages 467--477. IEEE, 2002.

\bibitem{Kuhn:2004}
D.~R. Kuhn, D.~R. Wallace, and A.~M. Gallo, Jr.
\newblock Software fault interactions and implications for software testing.
\newblock {\em IEEE Trans. Softw. Eng.}, 30(6):418--421, June 2004.

\bibitem{Meinicke:2016}
Jens Meinicke, Chu-Pan Wong, Christian K\"{a}stner, Thomas Th\"{u}m, and Gunter
  Saake.
\newblock On essential configuration complexity: Measuring interactions in
  highly-configurable systems.
\newblock In {\em Proceedings of the 31st IEEE/ACM International Conference on
  Automated Software Engineering}, ASE 2016, pages 483--494, New York, NY, USA,
  2016. ACM.

\bibitem{productlinesbook1}
Klaus Pohl, G\"{u}nter B\"{o}ckle, and Frank J. van~der Linden.
\newblock {\em Software Product Line Engineering: Foundations, Principles and
  Techniques}.
\newblock Springer-Verlag, Berlin, Heidelberg, 2005.

\bibitem{productline_survey}
Thomas Th\"{u}m, Sven Apel, Christian K\"{a}stner, Ina Schaefer, and Gunter
  Saake.
\newblock A classification and survey of analysis strategies for software
  product lines.
\newblock {\em ACM Comput. Surv.}, 47(1):6:1--6:45, June 2014.

\bibitem{static_slice}
Mark~David Weiser.
\newblock {\em Program Slices: Formal, Psychological, and Practical
  Investigations of an Automatic Program Abstraction Method}.
\newblock PhD thesis, Ann Arbor, MI, USA, 1979.
\newblock AAI8007856.

\bibitem{wong2009survey}
W.~E. Wong, R.~Gao, Y.~Li, R.~Abreu, and F.~Wotawa.
\newblock A survey on software fault localization.
\newblock {\em IEEE Transactions on Software Engineering}, 42(8):707--740, Aug
  2016.

\bibitem{Yin:2011}
Zuoning Yin, Xiao Ma, Jing Zheng, Yuanyuan Zhou, Lakshmi~N. Bairavasundaram,
  and Shankar Pasupathy.
\newblock An empirical study on configuration errors in commercial and open
  source systems.
\newblock In {\em Proceedings of the Twenty-Third ACM Symposium on Operating
  Systems Principles}, SOSP '11, pages 159--172, New York, NY, USA, 2011. ACM.

\end{thebibliography}
\end{document}